\documentclass[sigconf]{acmart}

\renewcommand\footnotetextcopyrightpermission[1]{} %
\usepackage{subcaption}

\AtBeginDocument{%
  \providecommand\BibTeX{{%
    \normalfont B\kern-0.5em{\scshape i\kern-0.25em b}\kern-0.8em\TeX}}}

\setcopyright{none}
\copyrightyear{2022}
\acmYear{2022}
\acmDOI{10.1145/1122445.1122456}

\acmConference{}{}{}
\acmBooktitle{}
\acmPrice{}
\acmISBN{}

\begin{document}

\title{Online Unplugged and Block-Based Cryptography in Grade 10}
\author{Michael Lodi}
\email{michael.lodi@unibo.it}
\orcid{0000-0002-3330-3089}
\affiliation{%
	\institution{Universit\`a di Bologna}
	\city{Bologna}
	\country{Italy}}
\affiliation{%
	\institution{INRIA Sophia-Antipolis}
	\city{Valbonne}
	\country{France}}

\author{Marco Sbaraglia}
\email{marco.sbaraglia@unibo.it}
\orcid{0000-0002-5623-4071}
\affiliation{%
	\institution{Universit\`a di Bologna}
	\city{Bologna}
	\country{Italy}}

\author{Simone Martini}
\email{simone.martini@unibo.it}
\orcid{0000-0002-9834-1940}

\affiliation{%
	\institution{Universit\`a di Bologna}
	\city{Bologna}
	\country{Italy}}
\affiliation{%
	\institution{INRIA Sophia-Antipolis}
	\city{Valbonne}
	\country{France}}

\renewcommand{\shortauthors}{Lodi et al.}

\begin{abstract}
	We report our experience of an extracurricular online intervention on cryptography principles in 10th grade. This paper's first goal is to present the learning path we designed, influenced by cryptography core ideas rather than technical knowledge. We will detail how we used Snap! (a visual programming language) to realize hands-on activities: programming playgrounds to experiment with cryptosystems and their limits, and interactive support for an unplugged activity on the Diffie-Hellman key exchange. The second goal is to evaluate our intervention in terms of both student perceptions and learning of core cryptography ideas. The students appreciated the course and felt that, despite being remote, it was fun, interesting, and engaging. They said the course helped them understand the role of cryptography, CS, and Math in society and sparked their interest, especially in cryptography and CS. The third goal is to discuss what worked well and areas of improvement. Pedagogically, remote teaching caused high ``instructor blindness'' and prevented us from giving the optimal amount of guidance during the exploration activities with Snap! playgrounds, making them sometimes too challenging for total programming novices. On the other hand, the ``remote-unplugged'' Diffie-Hellman worked well: it embodies a coherent metaphor that engaged the students and made them grasp this groundbreaking protocol. The students praised the activities as engaging, even when challenging. The final assessment showed that the core cryptography ideas were well understood.
\end{abstract}

%\ccsdesc[500]{Security and privacy~Cryptography}
%\ccsdesc[500]{Social and professional topics~K-12 education}
%\ccsdesc[300]{Security and privacy~Social aspects of security and privacy}

%\ccsdesc[300]{Applied computing~Distance learning}


\maketitle

\section{INTRODUCTION}

Cryptography is at the core of today's digital society's many activities and tools (e.g., instant messaging, e-commerce, stock exchange).
As recognized by~\cite{keepingsecrets}, various frameworks (e.g.,~\cite{digicomp}) and curricula (e.g., the K-12 Computer Science Framework~\cite{K12Framework} and the UK computing curriculum~\cite{ukcomputing}) include competencies related to cybersecurity.
Some of them are more oriented on using security for personal purposes, others on understanding how digital security works, but they all recognize that cybersecurity skills are essential for students to be active citizens of digital society.
The fact that cryptography is the foundation of cybersecurity implies its scientific and cultural value. In addition, novices identified cryptography ``as an interesting context for computer science lessons''~\cite[p. 3]{keepingsecrets}.
However, not much research has been done in teaching cryptography for K-12.
Considering that K-12 education does not aim to train professionals but to provide lenses to understand the world and skills to act in it, we consider it important to teach the fundamental principles of cryptography in a way that could be understood by novices.
We wanted to create a course that, while short and without prerequisites, would help students understand the principles of cryptography and their importance in our society. To do so, we designed a learning path and two non-traditional and complementary types of activity.
Due to the ongoing COVID-19 pandemic, we had to design the intervention as remote-only.

In general---since educational research has shown the effectiveness of active and cooperative learning methodologies~\cite[p. 304]{Loui2019}---we aimed to design meaningful, concrete, and interactive activities for students.
We developed cryptography playgrounds to let students know, use and attack simple cryptosystems (e.g., Ceasar cipher) and a ``remote-unplugged'' activity to make students (in pairs) enact the Diffie-Hellman key exchange. We realized both types of activities with Snap!, a visual
programming language.

Practical activities are steps of the learning path we developed around the fundamental cryptography principles. Our pathway includes a few meaningful systems and schemes while limiting technical knowledge. In particular, the need to overcome every time the limitations of these systems shapes a meaningful progression intended to support learning.

This paper's first goal is to present our learning path (section~\ref{sec:course}), focusing on the cryptography principles and schemes as motivations for content progression (\ref{ssec:path}) and the development and experience of Snap! activities---i.e., crypto playgrounds and unplugged Diffie-Hellman (\ref{ssec:tools}). The second goal is to evaluate our experience (sections~\ref{sec:exp} and~\ref{sec:discuss}) regarding both the level of understanding of the fundamental concepts covered (\ref{ssec:assessment_data}, \ref{ssec:disc_satis}) and students' satisfaction and perceived utility (\ref{ssec:satis}). The last goal is to discuss what worked and what did not (section~\ref{ssec:disc_satis}) and help CS educators adopt and adapt our pathway and practical activities (\ref{ssec:disc_adopt}).

\section{RELATED WORK}\label{sec:related}

In 2017, the Joint Task Force on Cybersecurity Education released the ``Curriculum Guidelines for Post-Secondary Degree Programs in Cybersecurity''~\cite{curricula2017}. Although they are targeted at post-secondary education, among the first concepts to be learned, they include ``basic concepts in cryptography to build the base for other sections in the knowledge
unit''~\cite[p. 24]{curricula2017}.
In the K-12 Computer Science Standards~\cite{CSTAStandard2017} proposed by U.S. Computer Science Teachers Association (CSTA) and ACM, cybersecurity is important for all grades. In particular, for Level 2 (Grades 6-8, 11-14 y.o.), the standard (2-NI-06) indicates that students should be able to
encode and decode messages with various encryption methods and understand their different levels of complexity, starting with the simplest (e.g., Caesar cipher) to the more complicated public-key ciphers, which are better learned through unplugged activities.

However, a review of cybersecurity education over the last ten years of SIGCSE and ITiCSE conferences~\cite{review10anni} found that research ``predominantly focus[es] on tertiary education in the USA''. In addition, very few papers specifically address cryptography.
Our search confirmed this.
While some research is focused specifically on university-level cryptography courses~\cite[see, e.g.,][]{Buchele2013, Hsin2005}, in many proposals, cryptography was just one of the many parts of a cybersecurity course or curriculum
\cite[see, e.g.,][]{Sommers2010,Turner2011,Brown2012,Deshpande2019},
most often applied as a tool rather than explained deeply in its inner workings.
The use of visualization tools or interactive simulations for cryptographic algorithms (e.g., Caesar, Vigenère, DES, AES, RSA, SHA) has been proposed~\cite{Simms2011} and developed~\cite{Schweitzer2009, Schweitzer2009labs, 10.1145/2899415.2899425}, showing how ciphers work, their weaknesses and possible attacks. Sometimes these visualizations and simulations are accurate but too detailed for young students.
Some ideas relevant to our work are: using a message board to simulate a person-in-the-middle attack on a public-key cryptosystem~\cite{Greenlaw2015};
having students implement cryptographic algorithms (from classical to modern) with computer algebra systems~\cite{McAndrew2008}; helping novice developers securely use cryptographic APIs by providing them with cryptographic building blocks designed in line with the visual metaphors of block-based programming languages~\cite{Linden2018}.

From an extensive search of various libraries, we found few papers specifically focused on teaching basic cryptography principles to K-12 students.
Again, many proposals are in the broader context of cybersecurity curricula, including some hands-on cryptography activities, often set in a motivating real-world context. For example: understanding a secure mail exchange with simulations of encryption, decryption, attacks on Caesar and Vigenère ciphers and the RSA cryptosystem~\cite{Gramm2012}; a social networking platform that includes a simple activity of implementing the Caesar cipher in Python~\cite{Zinkus2019}; summer camps to teach cybersecurity and computational thinking through robotics and block-based programming languages, with Caesar and Vigènere encryptions implemented in robots~\cite{Robots2019,Robots2020}.
Some unplugged activities have been proposed: the classical CS Unplugged Kid Crypto~\cite{BELL2003199}, which simulates the one-way function of public-key cryptography with a Perfect Dominating Set on a graph; activities on symmetric transposition ciphers in the context of a scout story~\cite{Perekietka2013}; a one-day workshop to learn about classical and modern cryptosystems with paper, pencil, and handmade materials~\cite{Konak2014}; unplugged activities on Caesar and Rail Fence ciphers for a classical Capture The Flag cybersecurity competition~\cite{Ford2017}; a physical simulation of the Diffie-Hellman exchange with colors with students impersonating Alice and Bob mixing food coloring~\cite{Fees_daRosa_Durkin_Murray_Moran_2018}.
In general, hands-on inquiry-based cryptography and cybersecurity activities have been claimed to improve K-12 students' self-efficacy and problem-solving skills~\cite{konak_experiential_2018}.

\section{OUR COURSE}\label{sec:course}

\subsection{Context}
In Italy, upper secondary school lasts five years (usually students start when they are 14 y.o.), %
but only the first two are compulsory.
Upper secondary level %
has a
a variety of strands: lyceums, tecnical and professional.
\textit{Lyceums} give a theoretical basis in classical, scientific or artistic areas and naturally lead to university studies.
See~\cite{bellettini_et_al_14} for a %
summary of the Italian secondary school system.

The \textit{Mathematical Lyceum}
is a national experimental project. It is an extra-curricular afternoon supplement available to all lyceum students on a voluntary basis. Students participate in extra laboratory activities to experience interdisciplinarity between Mathematics and other disciplines.
The project's goal is not to introduce more information but to reflect on ideas and foundations of knowledge and to broaden cultural horizons~\cite{liceomatematico}.
In this context, we accepted the invitation from a local lyceum to organize a cryptography course.

15 students (5 girls and 10 boys), all in their second year of lyceum (ca. 15-16 y.o.), participated in the course. None of them had previous programming experience.
The course was taught by two of the authors of this paper, who are researchers in computer science education and have experience in high school teaching.
We held four lessons (2h each) every Tuesday afternoon in February 2021.
Due to the ongoing COVID-19 pandemic, the lessons were held online through the Google Meet platform adopted by the school.

\subsection{Learning path}\label{ssec:path}

We designed the intervention with the dual-lens of \textit{seeing mathematical concepts applied to cryptography}, and more importantly, as computer scientists, \textit{teaching some fundamental principles of cryptography crucial in today's society}.

The development of our learning path was influenced by a broader research project~\cite{special} aimed at identifying the ``big ideas of cryptography'' (i.e., those core concepts, understandable by non-experts, that are ``beacons'' to guide teaching and learning) through interviews with experts and an open discussion with both educators and scholars.

\subsubsection{Contents}\label{subsub:contents}
\mbox{}

\emph{Day 1}
\begin{itemize}
	\item The social debate about encryption in
	      digital communication
	\item Caesar cipher: encoding, decoding, brute-force attack
	\item \textit{Homework: transposition vs. substitution, Kerckhoffs principle}
\end{itemize}

\emph{Day 2}
\begin{itemize}
	\item Caesar cipher: frequency attack
	\item One-time pad: encryption, decyrption, frequency attack
	\item \textit{Homework: encoding chars as bits, toy cryptosystem using XOR and bit swaps, hints at modern block ciphers (DES, AES)}
\end{itemize}

\emph{Day 3}
\begin{itemize}
	\item One-time pad: brute-force attack, perfect security, limitations, key-distribution problem
	\item Diffie-Hellman key exchange: simulation with colors
	\item \textit{Homework: quiz on One-time pad and Diffie-Hellman}
\end{itemize}
\emph{Day 4}
\begin{itemize}
	\item Math of Diffie-Hellman: modular arithmetic, exponential and its inverse, primes and coprimes (and hints at generators)
	\item Diffie-Hellman: key exchange, computational security, person-in-the-middle attack
	\item Asymmetric cryprography: terminology, key pairs properties, non-technical schemes for authentication and secrecy, intuitive idea of \textit{one-way function} (multiplying vs. factoring)
	\item Putting all together: how \textit{intuitively} combine asymmetric and symmetric schemes for authentication and secrecy
	\item \textit{Homework: satisfaction survey, fill-in-the-blanks assessment}
\end{itemize}

\subsubsection{Milestones and their motivation}

\begin{enumerate}
	\item \textbf{Caesar cipher}
	      \begin{itemize}
		      \item Motivations: basic example of sym-crypto, easy to understand and play with it
		      \item [$\boldsymbol{\downarrow}$] Problems to overcome: attackable with both brute-force and frequencies
	      \end{itemize}
	\item \textbf{One-time pad cipher}
	      \begin{itemize}
		      \item Motivations: resistant to both brute-force and frequencies attacks, implements perfect security
		      \item [$\boldsymbol{\downarrow}$] Problems to overcome: specific feasibility issues (and the more general key-distribution problem)
	      \end{itemize}
	\item \textbf{Modern symmetric cryptosystems}
	      \begin{itemize}
		      \item Motivations: fast to implement with modern computers (though ``only'' computationally secure)
		      \item [$\boldsymbol{\downarrow}$] Problems to overcome: the key-distribution problem
	      \end{itemize}
	\item \textbf{Diffie-Hellman key exchange}
	      \begin{itemize}
		      \item Motivations: creates a shared key on an insecure channel
		      \item [$\boldsymbol{\downarrow}$] Problems to overcome: person-in-the-middle attack
	      \end{itemize}
	\item \textbf{Asymmetric cryptosystems}
	      \begin{itemize}
		      \item Motivations: allow authentication and secrecy
		      \item [$\boldsymbol{\downarrow}$] Problems to overcome: computationally expensive
	      \end{itemize}
	\item \textbf{Hybrid cryptosystems}\footnote{We did not include problems here because hybrid cryptosystems were the final destination of our path.}
	      \begin{itemize}
		      \item Motivations: combine the best of both symmetric and asymmetric cryptosystems
	      \end{itemize}
\end{enumerate}

\subsection{Tools, activities and methodology}\label{ssec:tools}
\subsubsection{Tools used}\label{ssec:tools_used}
As a videoconferencing service, we used Google Meet. In general, we relied on \textit{Google Workspace for Education}, the platform chosen by the host school. More in detail, we used Google Classroom to share announcements, learning materials, and homework. We used Google Slides to present slides and simple animations of crypto schemes and Google Docs to share longer texts\footnote{E.g., homework readings from outreach materials, like Singh's ``The Code Book''~\cite{singh1999}.}.
Also, we used Google Forms to: i) guide free exploration activities with questions, ii) gather homework, and iii) collect final evaluation answers and students' feedback on the course.

Snap!~\cite{snap}
(originally BYOB, \textit{Build Your Own Blocks}) is a visual, block-based programming language. It is an extended reimplementation of Scratch with functions and many other features that make it suitable for a serious introduction to CS programming for high school or college students.
In particular, we choose Snap! because it allows creating new blocks (i.e., new functions). Unlike in Scratch, in Snap! custom blocks can return values and also hide their implementation (i.e., the function body) from students.

\subsubsection{Tools created, activities and methodology}\label{ssec:tools_created}

\begin{figure*}
	\centering
	\includegraphics[width=\textwidth]{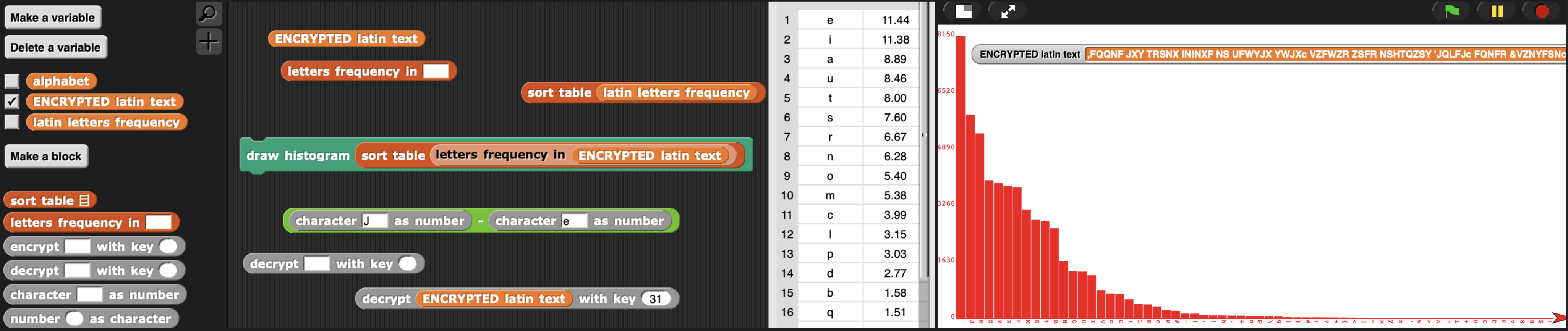}
	\caption{Attacking Caesar cipher with frequencies - Snap! playground}
	\label{fig:teaser}
\end{figure*}

Using Snap!, we created a progression of playgrounds to support the hands-on activities. These environments allow students to experiment with cryptosystems (e.g., Caesar cipher, One-time pad) and their limits (e.g., ease or difficulty of decryption, the computation time required). A playground is a Snap! project with a limited set of visible instructions. By leveraging Snap!'s ability to create custom blocks and hide existing ones, we provided only the blocks needed to encrypt and decrypt messages and carry out possible attacks for each cryptosystem.
For example, to attack a Caesar ciphertext using frequencies, some of the blocks available in the playground were: calculating frequencies in a text, sorting a matrix (i.e., letter and frequency) by frequencies, representing a two-dimensional matrix with an histogram, and a table of average frequencies in the used language (fig.~\ref{fig:teaser}).

All playgrounds used during the course---which follow the path outlined in~\ref{ssec:path}---are available for exploration and use~\cite{material}.
Through guiding questions and small coding, decoding, and attack challenges, we invited the students to play by combining the Snap! blocks. The explicit and shared objective was to solve the proposed task; the broader goal was for the students to experience the cryptosystem and understand its operation and limitations.
We always tried to design activities by contextualizing them in meaningful scenarios for students. For example, the activities on Caesar cipher involve a Latin\footnote{All our lyceum students were studying Latin.} text (to be used for a test) that a quarantined teacher wants to communicate secretly to her substitute colleague without the students being able to intercept it.

As the complexity of the systems increased, we felt that asking students to program the algorithms, albeit in the simplified way of blocks, was too great a challenge for the course, given that they were novices in programming.
For example, for the Diffie-Hellman key exchange, we thought that having an already working program that supported students in simulating the shared secret generation could suit better understanding the protocol at a high level.
So instead of providing a Snap! playground, we gave students an executable-only Snap! project (i.e., they could not modify it nor explore its code) that guided them in performing the actions of one of the two actors involved in a Diffie-Hellman key exchange~\cite{material}.
There are at least two typical ways to didactically present Diffie-Hellman~\cite[see, e.g.,][]{enwiki:1038627636}: one based on mixing colors (more evocative yet simplified), the other on performing the algorithm with small numbers (more accurate yet complex). We tried to combine these two modes to gradually move from color metaphor to mathematical functioning.
In particular, our Snap! program assigns a color to each number from $0$ to $99$.
Color mixing is not done with classical additive or subtractive algorithms but is based on the actual Diffie-Hellman calculations (initially hidden to the students), made between the simple numbers that in Snap! represent colors.
Through this program, the students in pairs were able to generate a shared secret color for every couple.
The necessary communications between the two students (i.e., choosing an initial shared color and exchanging the calculated color) took place on the  public chat, which effectively represented the insecure channel since everyone could listen over it.

After the students experienced firsthand the high-level functioning of the Diffie-Hellman algorithm, we explained the mathematical underpinnings necessary to understand its details (see~\ref{subsub:contents}).
At this point, we illustrated the algorithm itself through an animation~\cite{material}  that shows the correspondence between the colors and the simple numbers representing them, revealing the actual
calculations that drive the color mixing.
Then, we used other simple animations~\cite{material} of typical communication scenarios to illustrate increasingly complex concepts (e.g., secrecy and authentication asymmetric schemes) and abstract the technical
details to focus on understanding the underlying cryptographic principles.

\section{RELATIONSHIP WITH PREVIOUS WORK}

Our path is aligned with CSTA K-12 CS Standards~\cite{CSTAStandard2017}, both in terms of cryptography contents and methodologies (e.g., unplugged activities for public-key cryptosystems). It also covers almost all of the basic fundamental concepts of cryptography identified for tertiary education~\cite[pp. 24, 30]{curricula2017}, adapted to the age target.
Instead of seeking completeness and depth in technical details, our choice of topics was made to convey the fundamental ideas of cryptography.

In line with proposals by other authors (see section~\ref{sec:related}), our path is driven from start to finish by a motivating and timely social problem: end-to-end encryption in private messaging.
Classroom activities are contextualized in situations that are meaningful to students (e.g., secret exchange of an exam text).

Supported by the literature (sec.~\ref{sec:related}), we designed hands-on activities in which students can directly experiment with the workings and limitations of cryptosystems. We faced the additional challenge of teaching the course remotely and found a successful way to deliver an unplugged activity, taking advantage of the features of the new medium while maintaining the ``unplugged spirit'' (see~\ref{ssec:tools_created}).

Compared to other proposals (sec.~\ref{sec:related}), which often provide ready-to-use software visualizations or simulations, we took a first, important step in ``cracking the black box'':
we wanted to introduce some ``computational thinking'' by asking students to program parts of the experience (e.g., combining programming blocks to calculate and visualize letter frequencies in order to make a frequency attack).

\section{EXPERIENCE EVALUATION}
\label{sec:exp}

At the end of the course, we asked the students to fill two Google Forms to get their feedback and assess their learning.
No marks were foreseen for these final activities.
Both questionnaires were completed by the same 14 (out of 15) students.

\subsection{Learning assesment}\label{ssec:assessment_data}
\begin{figure}
	\centering
	\includegraphics[width=\columnwidth]{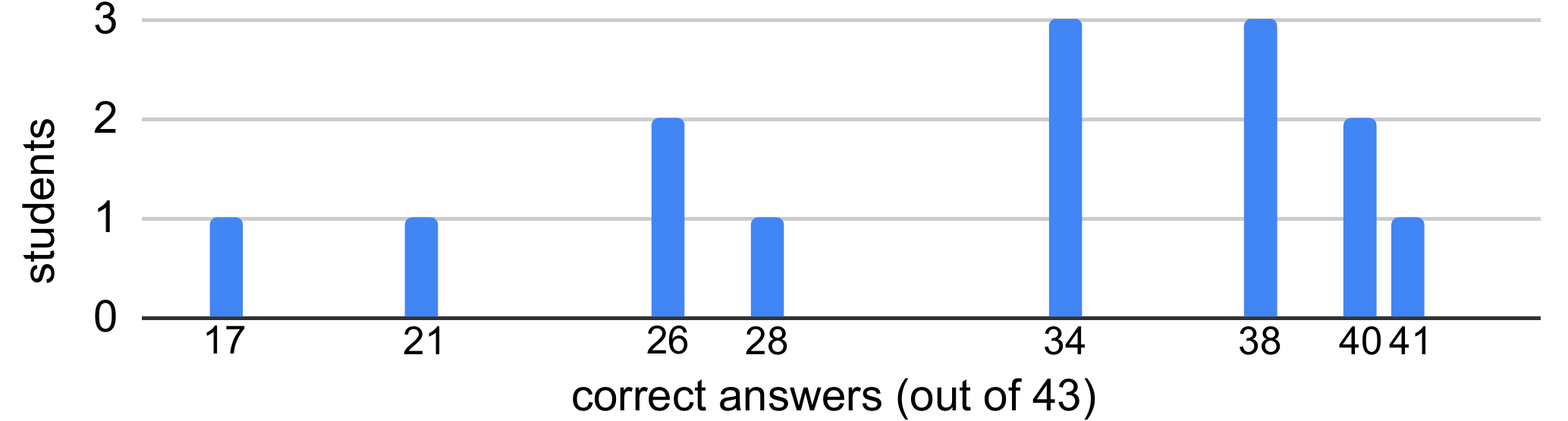}
	\caption{Distribution of correct answers}
	\label{fig:correct}
\end{figure}

We wrote a 2000-words summary of the important ideas and concepts covered in the course. In that summary, we identified 43 key passages. The students had to choose between the right filling and a wrong alternative we devised for each of them.
We wanted the activity also to be an opportunity for students to review the important contents of the course, so we structured it as a narrative\footnote{The english translation of the summary, with the right and wrong options, is available at~\cite{assessment}}.

The results were very positive: out of 43 questions, the mean of correct answers was 32.5, the median was 34, with a range of correct answers between 17 and 41.
See the distribution of correct answers per student in fig.~\ref{fig:correct}.

\subsection{Course satisfaction}\label{ssec:satis}

In this questionnaire, we asked students about their learning experience and the perspectives the course gave them. All answers were required.

Attendance was very high: out of 14, 13 students followed the first lesson, 12 the second, 11 the third, and 12 the fourth.

Regarding the course lenght, 9 students found it `Adequate', 4 `Too short', only 1 student found it `Too long'.

\begin{figure}
	\centering
	\includegraphics[width=0.95\columnwidth]{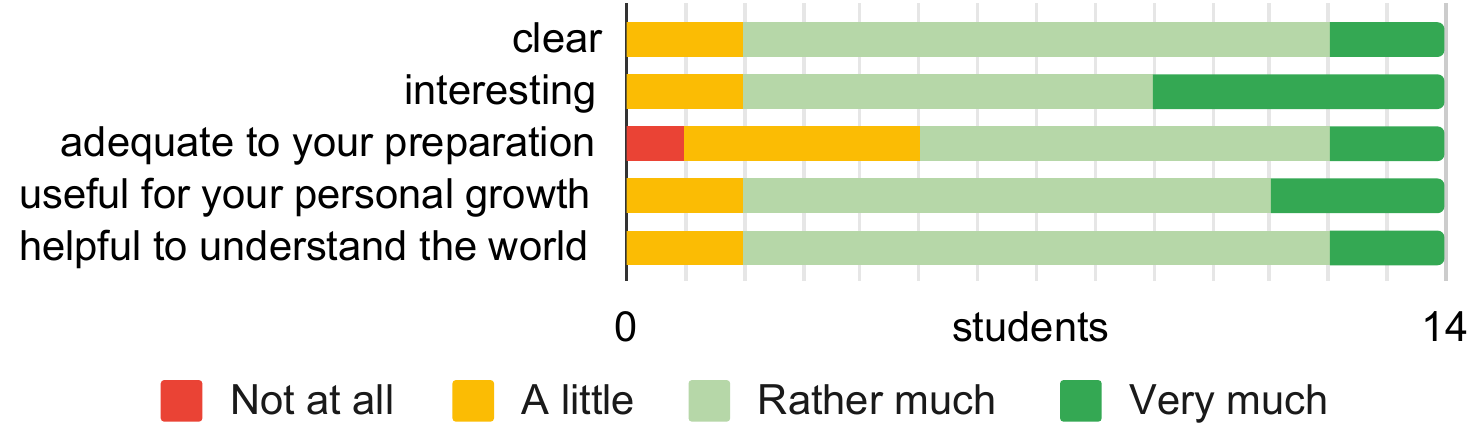}
	\caption{Student perceptions on course activities}
	\label{fig:courseactivities}
\end{figure}

About the activities, most of the students found them clear, interesting, useful for their preparation and for understanding the world. Some students felt that the activities were too difficult for their previous knowledge (fig.~\ref{fig:courseactivities}).

\begin{figure}
	\centering
	\includegraphics[width=0.9\columnwidth]{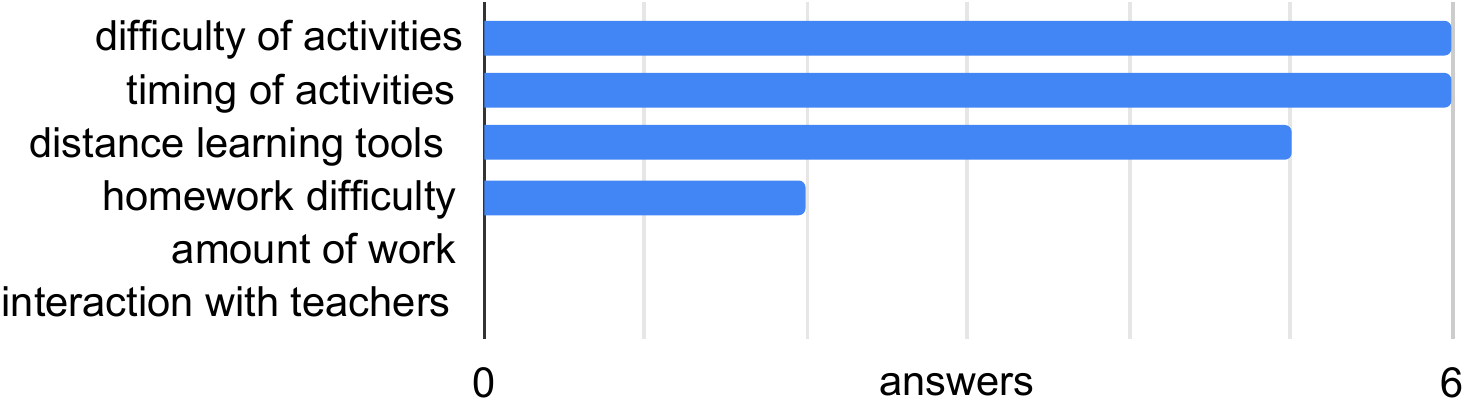}
	\caption{Student opinion on reasons of their difficulties}
	\label{fig:difficulties}
\end{figure}

When asked if they found difficulties during the course, on a four-point scale from `Never' to `Always', 1 students answered `Never', 11 students answered `Sometimes' and 2 students answered `Often'.
Most of the difficulties were related to the difficulty of the activities, the timing and organization, and the distance learning setting. See the details in fig.~\ref{fig:difficulties} (the same student could choose more than one option).

Regarding the satisfaction with the interaction with teachers and their support, on a four-point scale from `Not at all' to `Very much', 9 students answered `Rather much' and 5 `Very much'.

About the overall satisfaction with the course, on a four-point scale from `Not at all' to `Very much', 1 student answered `A little', 8 students `Rather much', and 5 students `Very much'.
13 students would suggest the course to a friend, 1 would not.

\begin{figure}
	\captionsetup[subfigure]{aboveskip=1pt}
	\begin{subfigure}[b]{\columnwidth}
		\centering
		\includegraphics[width=0.85\textwidth]{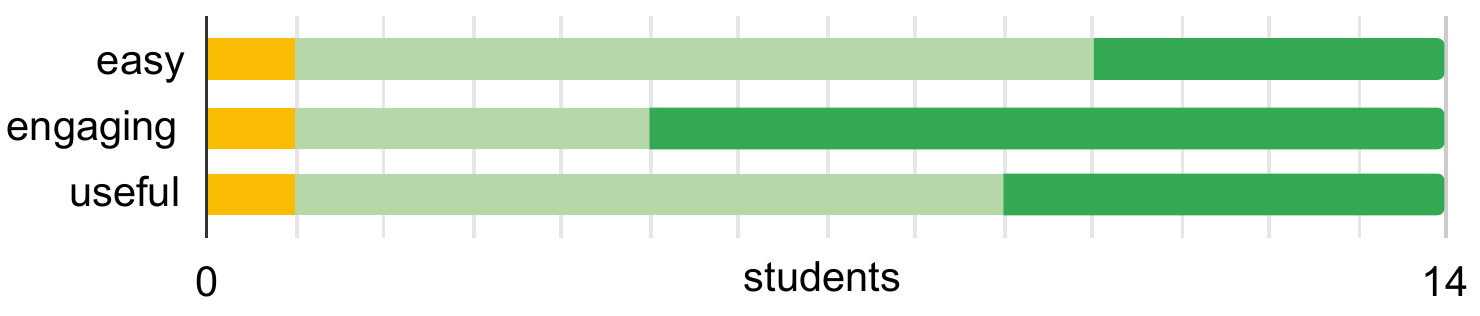}
		\caption{Diffie-Hellman color mixing activity}
		\label{fig:DHactivity}
	\end{subfigure}
	\par\medskip
	\begin{subfigure}[b]{\columnwidth}
		\centering
		\includegraphics[width=0.85\textwidth]{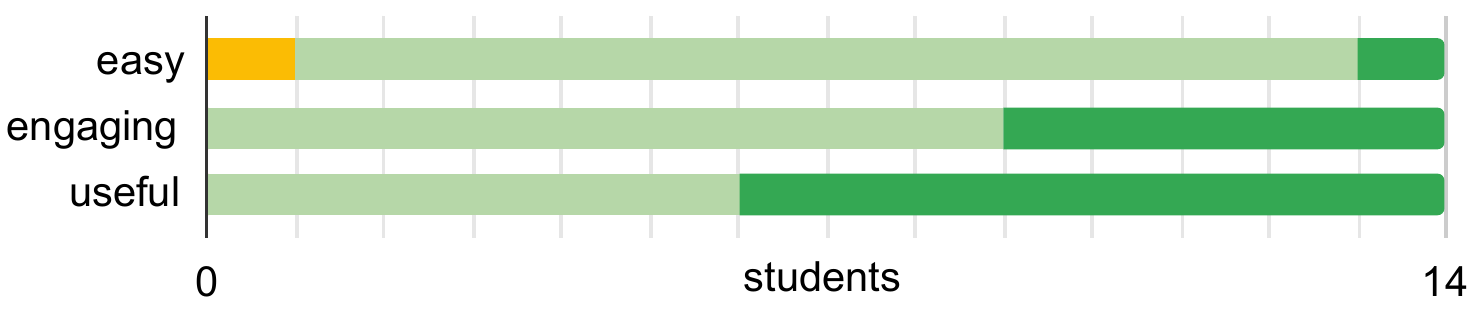}
		\caption{Explanations by animations}
		\label{fig:animations}
	\end{subfigure}
	\par\medskip
	\begin{subfigure}[b]{\columnwidth}
		\centering
		\includegraphics[width=0.85\textwidth]{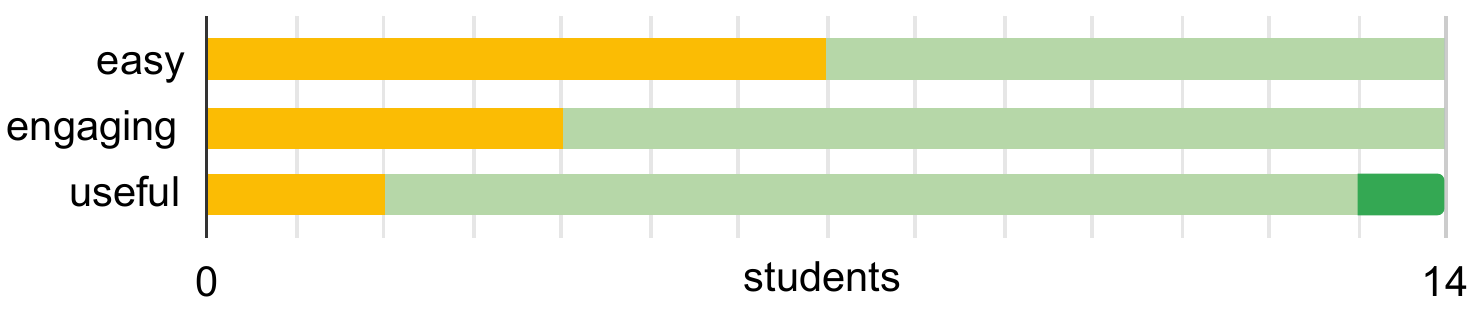}
		\caption{Home readings and reflection homework}
		\label{fig:homework}
	\end{subfigure}
	\par\medskip
	\begin{subfigure}[b]{\columnwidth}
		\centering
		\includegraphics[width=0.85\textwidth]{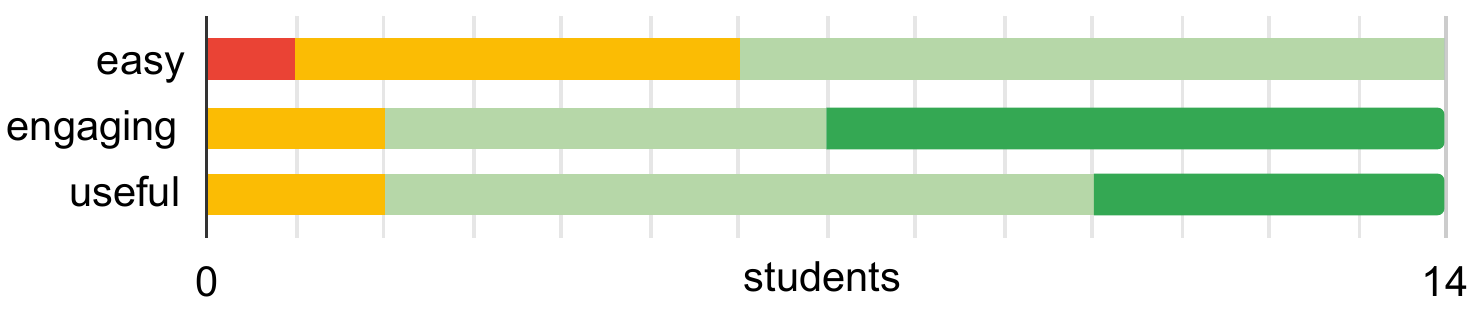}
		\caption{Snap! autonomous activities with blocks}
		\label{fig:snapactivities}
	\end{subfigure}
	\par\medskip
	\begin{subfigure}[b]{\columnwidth}
		\centering
		\includegraphics[width=0.85\textwidth]{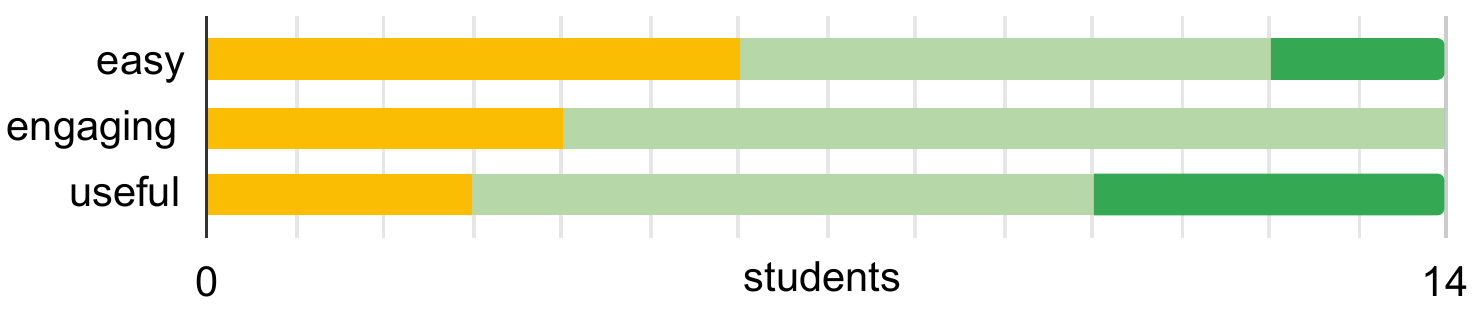}
		\caption{Using Google forms during activities}
		\label{fig:googleforms}
	\end{subfigure}
	\par\medskip
	\includegraphics[width=0.90\columnwidth]{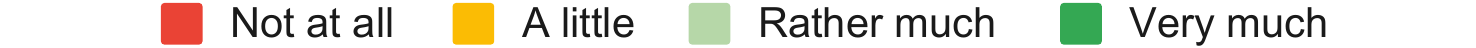}
	\caption{Student evaluation on course elements}
	\label{fig:allcourseelements}

\end{figure}

For what concerns tools used (see~\ref{ssec:tools_used}) and created (see~\ref{ssec:tools_created}), students highly appreciated the Diffie-Hellman activity (fig.~\ref{fig:DHactivity}) and the animated slide explainations (fig.~\ref{fig:animations}).
Some students found difficulties in both homework (fig.~\ref{fig:homework}) and in Snap! playgrounds (fig.~\ref{fig:snapactivities}).
Finally, some students found working with Google forms during lessons not very useful, engaging or easy (fig.~\ref{fig:googleforms}).

\begin{figure*}
	\centering
	\includegraphics[width=0.9\textwidth]{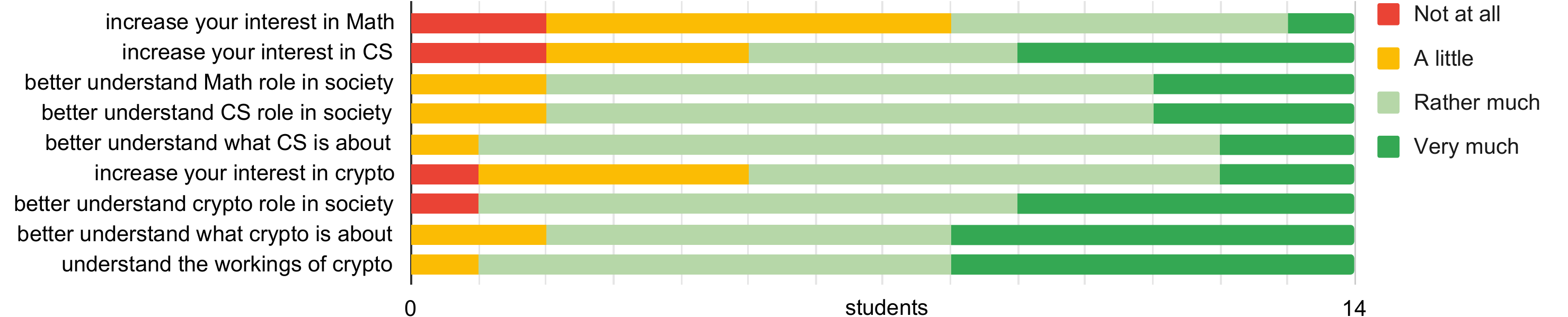}
	\caption{Student perceptions of what the course was useful for}
	\label{fig:finalusefulness}
\end{figure*}

The perceived utility of the course is relevant: most of the students found the course useful to better understand cryptography, its matter of study, its role in society.
Students also perceived to have better understood CS and Math role in society.
While the course clearly sparked interest about cryptography in most of the students, it also helped stimulating interest in CS in 2/3 of them, while less than half of them felt it has increased their interest in Math.
For a more detailed picture, see fig.~\ref{fig:finalusefulness}.

\subsubsection{Open comments from students}

All but one of the comments were positive.
Most students remarked that the activities were interesting and fun (e.g., ``I liked the fact that through Snap! we were able to play and experiment with cryptography'').
Comments described the lessons as well organized and engaging, even remotely.
One student said she really appreciated the space reserved for collective discussions in each lesson.

Many students would have liked the course to be longer.
One student would have wanted to delve into Snap! programming beyond the scope and boundaries of the proposed playgrounds.

The only non-positive comment reports difficulties in understanding and suggests more explanation, also through animations (highly appreciated, fig.~\ref{fig:animations}), before the hands-on activities.

As a positive note, a student used the Caesar cipher playground to encrypt (with a key we had to figure out) her positive comment.

\section{OBSERVATIONS AND FINDINGS}\label{sec:discuss}

We summarize the observations we gathered as educators and the findings of the assessments. We also provide CS educators with suggestions for adoption and possible extensions.
\subsection{On pedagogy and intervention results}\label{ssec:disc_satis}

The ``remote-unplugged'' Diffie-Hellman activity was received very well: simulating the insecure channel through the meeting public chat was an essential metaphor to understand the key-exchange.
In their answers, students said that the activity was engaging and fun, confirming our positive impressions on the spot.
We refer to it as ``remote-unplugged'' because it has almost all the characteristics of a CS Unplugged activity (i.e., \textit{real computer science}---presenting fundamental CS concepts and algorithms; \textit{learning by doing}; \textit{fun}; \textit{co-operative}; \textit{stand-alone}; \textit{resilient} to student errors; see~\cite{CSUPrinciples}) except that it was delivered via technological devices. In this case, however, the devices acted as a necessary means of communication rather than being the specific tool of computer scientists.
In addition, we believe that an interactive, executable-only Snap! project that guided the students through the steps of the Diffie-Hellman activity made the activity more concrete and easier to follow than with pen and paper and a list of rules, especially in a remote setting.

The Snap! playgrounds we carefully designed and implemented to let students play with cryptosystems through limited sets of custom primitives (e.g., encoding, decoding, attacking) were praised as engaging and useful.
For the easier concepts (e.g., Caesar cipher), the block-based programming playgrounds worked well as a way to get hands-on experience and understand the elements of a cryptosystem, some of its possible attacks, and the main limits (e.g., the computational time required).
However, half of the students found the more advanced playgrounds difficult.
We believe that the setting did not help: the course was held online, with students we did not know in advance (and mostly did not know each other, coming from different classes).
This resulted in severe ``instructor blindness,'' making it hard for us to act as facilitators and provide the students exploring the Snap! playgrounds with the \textit{optimal} amount of guidance~\cite{TaberOptimally}. Consequently, these activities were ``minimally guided''~\cite{Tobias_2009}, which can be too hard, especially for weaker students.

Hence, our aim to ``teach'' at least the fundamental principles of the path devised (see~\ref{ssec:path}), even in the short duration of the course, led us to a less open approach. First of all, we moved to a more guided exploration with the ``remote-unplugged'' activity on the Diffie-Hellman algorithm.
After that, in the final phase, we moved to a more traditional approach to illustrate the more advanced schemes of asymmetric cryptography, yet maintaining a Socratic approach with discussions and suggestions from the students. We taught the fundamental mechanisms behind secrecy and authentication through engaging animations~\cite{material} (using Power Rangers so that actions and messages of the different actors would be evident through their respective colors), leaving out technical and implementation details. Students highly appreciated the animations, explicitly saying they were really helpful to understand.

The final assessment focused on cryptography core ideas, in line with the goals of our intervention. It showed very good results (see~\ref{ssec:assessment_data}), indicating that the main contents of the course were received.
This is good for two reasons: because every citizen should understand today's digital society and because it is an opportunity for orientation towards the disciplines involved.
The satisfaction survey confirmed the achievement of these two goals, indicating that the students better understood the cryptography role in society, its matter of study, and increased their interest in it.

\subsection{Suggestions for adoption and adaption}\label{ssec:disc_adopt}

To help other educators adopt and adapt our course, we provide the lessons' contents and the related conceptual milestones (see~\ref{ssec:path}), animations and Snap! projects~\cite{material},
and the final assessment text~\cite{assessment}.
From a methodological point of view, we suggest giving more time (ideally double) for the Snap! programming playground activities to let all students get used to the environment and construct their knowledge by hands-on exploration.
The level of guidance for these activities can be adjusted. If the course is in presence, the teacher (with the help of teaching tutors if the number of students is high) can get a clearer sense of the students' difficulties and address them immediately while still leaving a high degree of freedom. If online, where it is harder to have the pulse of the students' needs, intermediate steps between the playgrounds can be considered, with more frequent check-ins and moments of realignment.

Additional time can also be foreseen to address other important topics (e.g., binary representation and operations as underlying mechanisms of modern symmetric block ciphers) during lessons rather than just on homework so that students can benefit from teacher support in the early stages of learning.

Having even more time, learning also some of the technical aspects (e.g., definitions of cryptosystem elements, representation of and operations on binary numbers, cryptography-related math topics, mathematical aspects of cryptographic algorithms) could be another explicit goal of this kind of intervention. If so, formative and final assessments also need to consider these specific learning objectives.

Looking ahead, our approach has the inherent potential to pursue learning goals related to computational thinking.
While the Snap! playgrounds we developed have narrow scopes on specific cryptosystems, the proposed activities involve some programming concepts (e.g., sequence, function composition, variables, lists).
More custom blocks can be developed, as well as activities that require a more extensive use of programming (e.g., using other fundamental elements of structured programming like conditionals and loops; ``looking inside'' custom Snap! blocks to adapt them to specific needs).
From a learning programming perspective, however, inspecting the code of our Snap! playgrounds does not have the straightforward educational value it could have. The student who had the curiosity would find extensive use of Javascript needed to integrate the functionalities of Snap! and some uninteresting workarounds put in place to overcome its limitations. Therefore, we plan to build a stack of \textit{notional machines} at different levels of abstractions so that students can progressively see more details by looking inside the blocks without being overwhelmed by all the complexity at once~\cite{sbaraglia-notional}.

\bibliographystyle{ACM-Reference-Format}
\bibliography{sample-base}

\end{document}